\documentclass[10pt,conference]{IEEEtran} %% for IEEE-CSR
\IEEEoverridecommandlockouts
\usepackage{cite}
\usepackage{amsmath,amssymb,amsfonts}
\usepackage{graphicx}
\usepackage{textcomp}
\usepackage{xcolor}

\IEEEoverridecommandlockouts
\def\BibTeX{{\rm B\kern-.05em{\sc i\kern-.025em b}\kern-.08em
    T\kern-.1667em\lower.7ex\hbox{E}\kern-.125emX}}
\begin{document}

\title{Machine Learning-based Ransomware Detection\\
Using Low-level Memory Access Patterns \\
Obtained From Live-forensic Hypervisor}

\author{\IEEEauthorblockN{Manabu Hirano}
\IEEEauthorblockA{\textit{Department of Information and Computer Engineering} \\
\textit{National Institute of Technology, Toyota College}\\
Toyota, Japan \\
hirano@toyota-ct.ac.jp
}
\and
\IEEEauthorblockN{Ryotaro Kobayashi}
\IEEEauthorblockA{\textit{Faculty of Informatics} \\
\textit{Kogakuin University}\\
Tokyo, Japan \\
ryo.kobayashi@cc.kogakuin.ac.jp}
} % author

\maketitle

%% not exceeding 150 words in submission system (147 words)
\begin{abstract}
Since modern anti-virus software mainly depends on a signature-based static analysis, they are not suitable for coping with the rapid increase in malware variants. Moreover, even worse, many vulnerabilities of operating systems enable attackers to evade such protection mechanisms. We, therefore, developed a thin and lightweight live-forensic hypervisor to create an additional protection layer under a conventional protection layer of operating systems with supporting ransomware detection using dynamic behavioral features. The developed live-forensic hypervisor collects low-level memory access patterns instead of high-level information such as process IDs and API calls that modern Virtual Machine Introspection techniques have employed. We then created the low-level memory access patterns dataset of three ransomware samples, one wiper malware sample, and four benign applications. We confirmed that our best machine learning classifier using only low-level memory access patterns achieved an $F_1$ score of 0.95 in detecting ransomware and wiper malware.
\end{abstract}


\section{Introduction}

%% Problems with variants and signature-based detection method
Ransomware is a type of malware that encrypts or exfiltrates victims' files to demand ransom. Many cyber-criminals create new ransomware variants using Ransomware-as-a-Service (RaaS) and ransomware toolkit, and they can easily evade protections of anti-virus software using the built-in obfuscators and packers \cite{sharmeen2020avoiding}. Although anti-virus software vendors frequently update their signature database used to detect binary files of the variants, the signature-based static analysis detection fundamentally cannot cope with a large number of the variants. Beaman et al. presented a literature review on recent state-of-the-art ransomware prevention and detection approaches \cite{beaman2021ransomware}. They analyzed popular ransomware samples and developed their experimental ransomware, AESthetic, that was able to evade detection against eight popular anti-virus programs. Their experiment highlights that current anti-virus software relies heavily on signature-based static analysis detection. They claimed that vendors should invest more into the approaches seen in recent literature, especially in dynamic analysis using behavioral patterns of ransomware.

%% Problem of vulnerable OS-based protection and mitigation using hypervisor (Microsoft VBS) 
Another problem of the current cyber-defense landscape is the evasion techniques of the conventional protection mechanisms. Many vulnerabilities of the complex software of modern OSs enable cyber-criminals to evade the detection mechanisms. Therefore, a more resilient multiple-layered ransomware detection mechanism is needed to address the advanced evasion techniques of the state-of-the-art malware.
Microsoft, for example, has introduced Virtualization Based Security, or VBS, in the latest products beginning with Windows 10 and Windows Server 2016 \cite{2021windowsinternals}. Microsoft's VBS uses the Windows hypervisor to create an additional protection layer under the conventional OS-based protection layer. Even when attackers gain access to the OS kernel, the extra protection layer limits the possible exploits. The Hypervisor-protected Code Integrity (HVCI) function of VBS, for example, can check whether the attacker modified the OS kernel and device driver code or not. Even if attackers evaded the first protection layer of an operating system (e.g., anti-virus, sandbox, and firewall), the following protection layer of a hypervisor (e.g., Microsoft's VBS, a hypervisor-based monitoring system presented in this paper) can detect the advanced cyber attacks.

\subsection{Literature review}
\label{sec:previous}

%% The relationship between this paper and the author's previous papers
Shinagawa et al. released an open-source lightweight hypervisor named BitVisor that transparently enforces security functions to a guest OS \cite{bitvisor}. Although Microsoft's VBS employs a multiple guest OSs model and enforces security functions from a privileged guest OS to a guest OS, BitVisor employs a single guest OS model and enforces security functions directly from the hypervisor to a guest OS. The single guest OS model of BitVisor reduces the performance overhead of a hypervisor-based security mechanism. In our previous work\cite{hirano2017waybackvisor}, we presented a BitVisor-based live-forensic hypervisor for collecting low-level storage access patterns. The low-level storage access patterns, the behavioral features obtained from the hypervisor, can be used to classify ransomware samples from benign applications using conventional machine learning models \cite{Hirano2019}. The low-level behavioral features of storage access patterns consist of variance of the accessed sector's logical block address, the entropy of written sectors, and throughput on a serial ATA (Advanced Technology Attachment) storage device (e.g., Solid State Drives, SSDs). The proposed system achieved the average $F_1$ score of 96.2\% in detecting ransomware and 94.1\% in detecting ransomware variants. Furthermore, we have released the open dataset of the low-level ransomware storage access patterns for researchers who are interested in constructing machine learning models for ransomware detection \cite{hirano2022ransap}. In this paper, we added a novel monitoring function of memory access patterns to the live-forensic hypervisor presented in the previous papers \cite{hirano2017waybackvisor,Hirano2019,hirano2022ransap}. 

%% moved from the Discussion section

Kumara and Jaidhar presented the Virtual Machine Introspection (VMI) technique that gathers details about the running processes of malware by introspecting the semantic view of a guest OS \cite{AJAYKUMARA201799}. Cheng et al. presented a lightweight live memory forensic framework based on hardware virtualization \cite{CHENG201723}. They showed a forensic method to obtain accurate information on malware processes using the information inside and outside the operating system. They also presented a forensic technique to monitor memory modifications by controlling page-grained permissions using EPT violations. Many state-of-the-art papers, including Sharmeen et al. \cite{sharmeen2020avoiding} and Leon et al. \cite{leon2021hypervisor}, have employed process-level dynamic features. While these papers \cite{CHENG201723, AJAYKUMARA201799, sharmeen2020avoiding, leon2021hypervisor} show a malware analysis method using process-level information, this paper presented a ransomware detection method using only low-level memory access patterns without bridging semantic gap. Our approach is also different from other hypervisors that mainly focus on dumping memory contents \cite{vis, hypersleuth}. 

\subsection{Contributions of this paper}
\label{sec:contribution}

This paper presents a novel machine learning-based ransomware detection method using low-level memory access patterns. The contributions of this paper are as follows.

\begin{itemize}
    \item  We enhanced the functionality of the BitVisor, a thin and lightweight hypervisor, by adding a function to collect low-level memory access patterns with mitigating advanced evasion techniques of malware that exploit OS vulnerabilities.
    \item While many modern ransomware detection methods presented in literature use dynamic features obtained from an operating system layer \cite{beaman2021ransomware} (e.g., sequences of API calls per process ID, file system operations), our system uses only low-level memory access patterns of physical address space obtained from a hypervisor layer. Many researchers believe that Virtual Machine Introspection (VMI), in general, needs to bridge the semantic gap between OS and hypervisor \cite{more2014virtual}. For example, low-level memory access patterns on RAM alone cannot determine which process or who accessed the memory region. This paper examines how we can use low-level memory access patterns obtained from a hypervisor layer to discriminate ransomware from benign applications without solving the semantic gap problem.
\end{itemize}

%%%%%%%%%%%%%%%%%%%%%%%%
\section{Live-forensic hypervisor for collecting memory access patterns}
\label{sec:surveillance}

%% Memory virtualization and SLAT
We first define the memory access patterns and feature vectors for creating machine learning models. Before describing them, we need to understand how hypervisors see memory address space of RAM. When virtual machines are executed on a machine, the hypervisor needs to translate a guest OS's physical address to a host machine's physical address. The memory address translation thus needs to be performed twice, once inside the guest OS from a guest virtual address to a guest physical address, and once inside the hypervisor from a guest physical address to a host physical address. Many modern hypervisor programs employ Second Level Address Translation (SLAT), a hardware-assisted address translation technology for OS virtualization, such as Intel's Extended Page Table (EPT) \cite{intel-sdm} and AMD's Rapid Virtualization Indexing (RVI) to improve the performance of the latter translation.

%% How EPT works
Intel's EPT works as follows: when a guest OS accesses a memory page of 4KB or 2MB for the first time, a VM exit due to an EPT violation (i.e., a hypervisor's page fault) activates a hypervisor. The hypervisor then creates a new EPT entry to map the guest physical address to a host physical address, followed by the guest OS's address translation to map the guest virtual address to the host physical address. Thus, an EPT violation occurs when there is no EPT entry for a memory page. Another reason for EPT violations is unauthorized access to the pages. An EPT entry of each host physical address has privilege bits of read, write, and instruction fetch (i.e., execute) to prohibit access on the memory page. The privilege bits protect a critical memory region (e.g., kernel code and drivers) from attackers on a guest OS. When a guest OS accesses a memory page that is not permitted in the page's EPT entry, a VM exit occurs due to an EPT violation; the hypervisor then decides whether the guest OS should execute the memory operations or not. 

\begin{figure}[tb]
    \centering
    \includegraphics[scale=0.38]{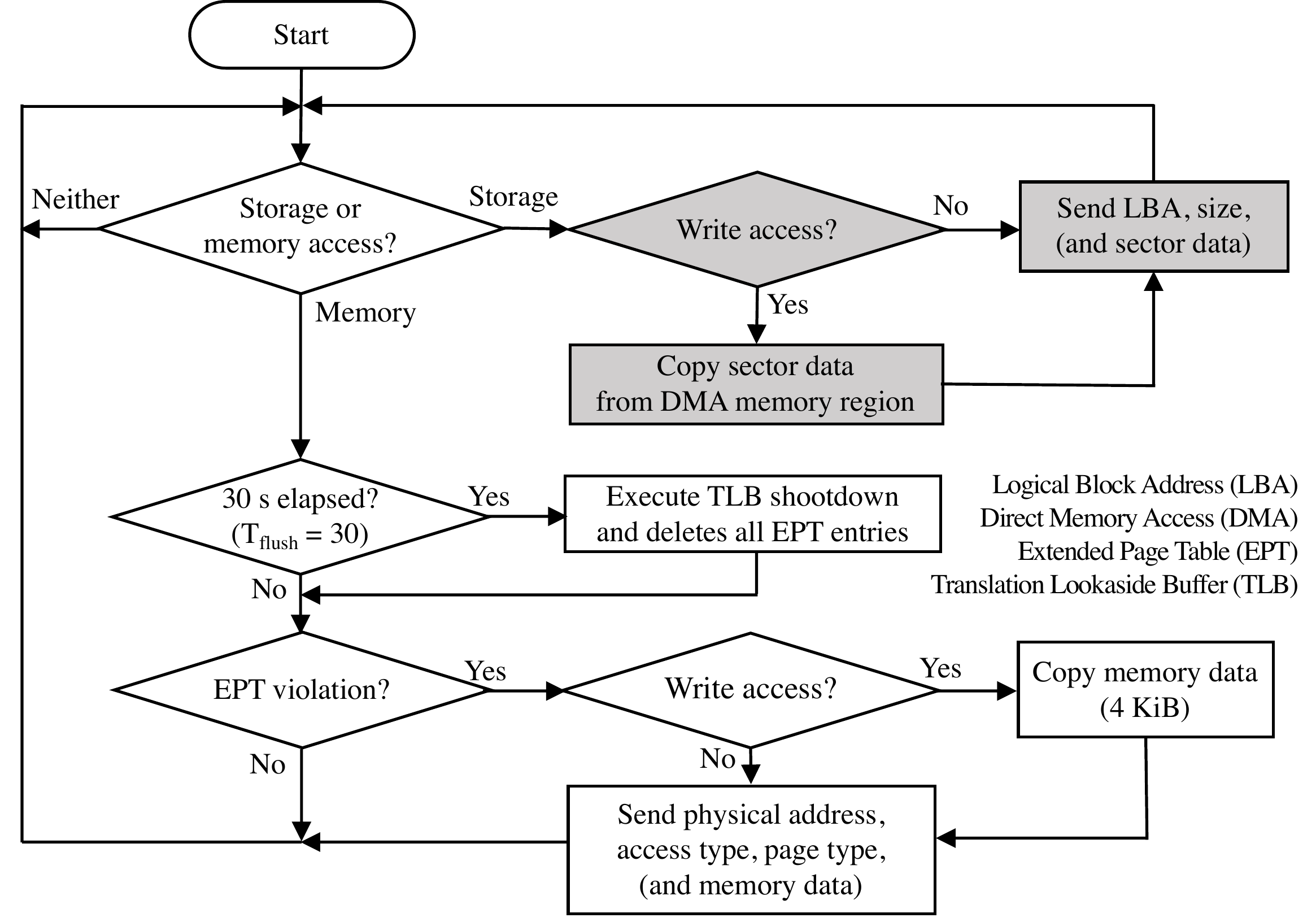}
    \caption{Flow chart of the live-forensic hypervisor that collects memory and storage access patterns.}
    \label{fig:flow-surveillance}
\end{figure}

%% Detail of the flow chart; how we monitor memory accesses using EPT violations.
Fig. \ref{fig:flow-surveillance} shows the flow chart of the live-forensic hypervisor that collects memory and storage access patterns. Although many Virtual Machine Introspection (VMI) systems drop specific permission on the particular EPT entries to deliberately cause EPT violations for intercepting memory accesses, the hypervisor presented in this paper deletes all EPT entries and executes a Translation Lookaside Buffer (TLB) shootdown every 30 s for intercepting memory accesses. TLB shootdown is an operation that flushes a TLB, a cache of address translation, on all CPU cores. The TLB shootdown on BitVisor was implemented by Fukai et al. \cite{Fukai}; we used their implementation as it is. The appropriate interval is discussed in Section \ref{sec:interval}.

When an application or a kernel program on a guest OS accesses the pages that are not both in TLBs and EPT entries, a VM exit due to an EPT violation activates the hypervisor. The hypervisor then reads Exit Qualification to obtain the access type (i.e., read, write, or instruction fetch). It also obtains the page type of the accessed page, including 4KB page, 2MB page, and Memory Mapped Input and Output (MMIO). MMIO is a method of performing input and output between CPU and peripheral devices using memory address space. When the access type is write, and the page type is not MMIO, the developed hypervisor obtains the memory contents at the physical address after executing five main loops of the hypervisor to ensure that the memory contents are updated from the original one. The obtained memory contents are mainly used to calculate entropy in feature engineering. The current implementation of the hypervisor collects only the first 4,096 bytes at the physical address even when a 2MB page was accessed and skips a byte filled with zeros. Please note that the design and implementation of the monitoring function of storage access patterns, which are shown in the grey symbols, were presented in the previous paper \cite{Hirano2019}.

%%%%%%%%%%%%%%%%%%%%%%%%
\section{Design and implementation}

\begin{figure*}[tb]
    \centering
    \includegraphics[scale=0.46]{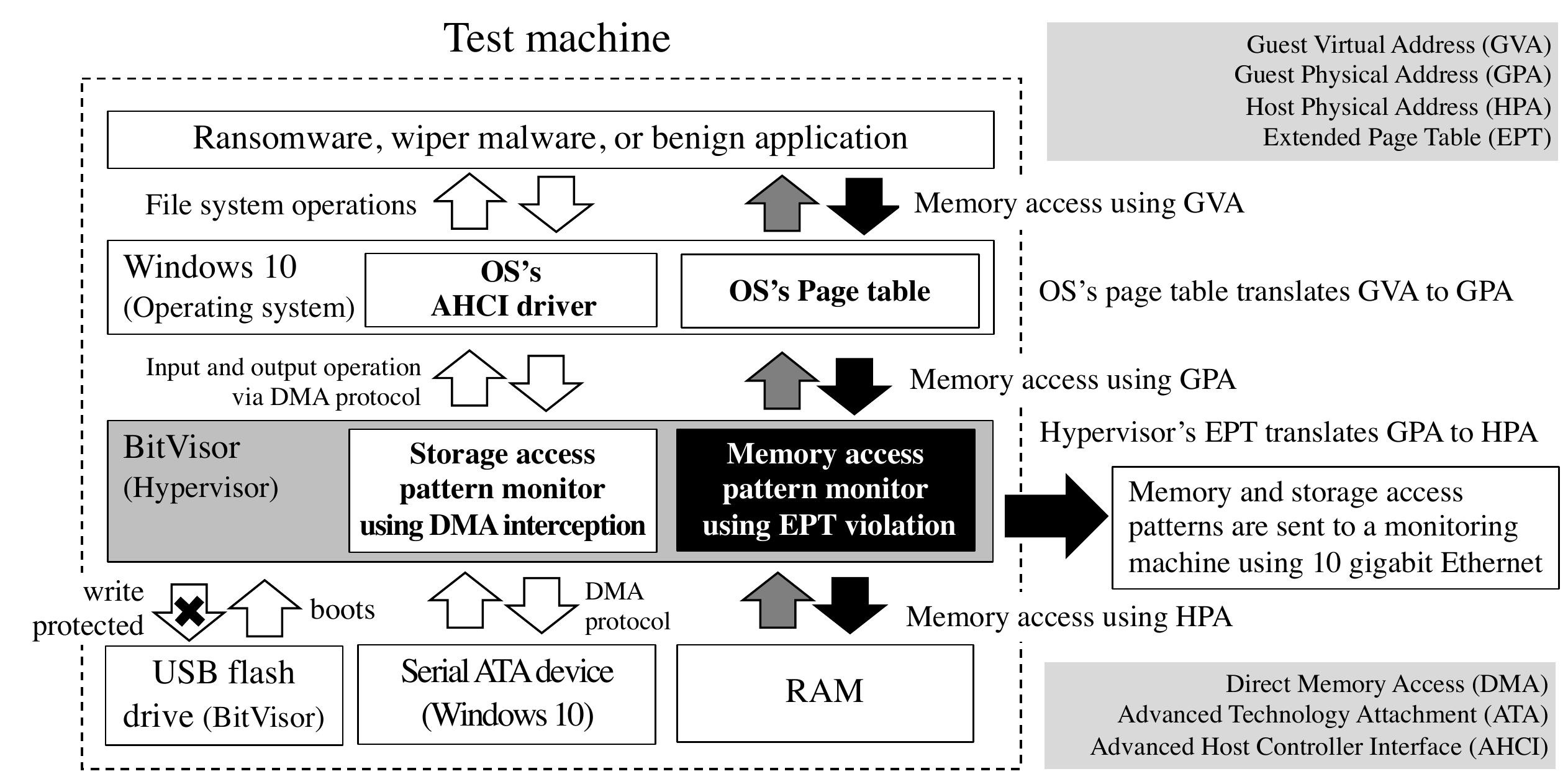}
    \caption{Schematic diagram of the live-forensic hypervisor for collecting memory and storage access patterns.}
    \label{fig:setup-procedure}
\end{figure*}

%% Schematic diagram
Fig. \ref{fig:setup-procedure} shows the schematic diagram of the hypervisor that collects memory and storage access patterns. Both the memory and storage access pattern monitor were implemented with BitVisor \cite{bitvisor}, a thin hypervisor for enforcing security functions using a single guest OS model. The security functions of BitVisor are transparently executed to a guest OS, and hence any modification of guest OSs and applications are not needed.
Table \ref{tbl:spec-test-machine} shows the specification of the test machine that executes the developed hypervisor. While a guest OS (i.e., Windows 10) uses a GbE card, the hypervisor uses a 10GbE card to transfer memory and storage access patterns to another monitoring machine. The SSDs were formatted in New Technology File System (NTFS) with GUID Partition Tables (GPT).

%% Workflow
The hypervisor is booted from a write-protected USB thumb drive to prevent overwriting our trusted computing base, the hypervisor program, from ransomware and wiper malware. The guest OS is then booted from an SSD, a serial Advanced Technology Attachment (ATA) device. The hypervisor monitors memory access patterns using EPT violations as described in Section \ref{sec:surveillance}. Since BitVisor employs a single guest OS model, the guest physical addresses and the host physical addresses are identical except for the memory region that BitVisor uses. The hypervisor also monitors storage access patterns by intercepting the Direct Memory Access (DMA) protocol \cite{Hirano2019}. 
The memory and storage access patterns are sent to another machine using User Datagram Protocol (UDP) packets on a 10-gigabit Ethernet connection with jumbo frames of 9,000 bytes. We can therefore collect access patterns even when a ransomware or wiper malware sample encrypted or destroyed data in serial ATA devices (e.g., SSD). BitVisor does not use virtual disk files but uses a serial ATA device as it is. We employ a drive duplicator to overwrite the entire contents of a serial ATA device with the initial content before executing a malware sample.

\begin{table}[tb]
\begin{center}
\caption{Specification of the test machine that executes the live-forensic hypervisor.}
\label{tbl:spec-test-machine}
\begin{tabular}{|l|l|}
\hline
CPU & Intel Cerelon G3920 (2.9 GHz) \\
\hline
RAM & DDR4-2133 8 GiB \\
\hline
Motherboard & ASRock H110M-HDV \\
\hline
Network & Intel Pro1000 (GbE), Intel X550-T1 (10GbE) \\
\hline
SSD & Crucial SSD 250GB (CT250MX500SSD1) \\
    & SSD Samsung SSD 250GB (MZ-7TD250) \\
\hline
OS  &  Windows 10 Enterprise LTSC (10.0.17763.2183) \\
\hline
Hypervisor & BitVisor downloaded on 24th Dec. 2021 (8c129a1) \\
\hline
\end{tabular}
\end{center}
\end{table}

%%%%%%%%%%%%%%%%%%%%%%%%
\section{Feature engineering}

We collected memory access patterns consisting of the four access types: read, write, execute, and read/write. The access type is obtained from Exit Qualification of a VM exit due to an EPT violation; when both read and write bits are 1, we process the access type as read/write. The following five feature vectors are calculated in each access type:

\begin{itemize}
    \item{Shannon entropy of a written page, $H(t)$ }
    \item{The number of accessed 4KB pages, $C_{4KB}(t)$ }
    \item{The number of accessed 2MB pages, $C_{2MB}(t)$ }
    \item{The number of accessed MMIO pages, $C_{MMIO}(t)$ }
    \item{Variance of accessed physcal addresses, $V(t)$ }
\end{itemize}

where $t$ is the time elapsed after executing a sample. A Shannon entropy $H(t)$ is calculated only when the access type is write or read/write. We, therefore, create 18-dimensional feature vectors (i.e., the five feature vectors for write and read/write, and the four feature vectors for read and execute) to make machine learning models.

Shannon entropy is a metric to measure uncertainty in a series of bit patterns \cite{6773024}. Ransomware encrypts many files, and the encryption operations increase the entropy of a specific memory region of RAM; hence, Shannon entropy can be a good indicator of ransomware activities. $H(addr)$ is a Shannon entropy of a written memory page at a host physical address $addr$, and it is calculated as follows:

\begin{equation}
\label{eq:entropy}
H(addr) = - \sum_{i=1}^{n} p(x_i) \log_2{p(x_i)}
\end{equation}

where $ p(x_i) $ is a probability of a byte $x_i$, which is an $i$th byte beginning from a physical address $addr$, and $n$ is the byte size of the collected memory data. Since the hypervisor cannot transfer all the contents of a 2MB page with acceptable performance, we employed 4,096 for $n$ in creating the dataset. $H(addr)$ produces a value between 0 and 8, where 8 represents a perfectly even distribution of byte values, where 0 represents a sequence of the same byte values. Before training machine learning models, the Shannon entropy is normalized between 0 and 1. A Shannon entropy at $t$, $H(t)$, is calculated as follows:

\begin{equation}
    H(t) = \frac{1}{N} \sum_{i=1}^{N} H_i(addr)
\end{equation}

where $H_i(addr)$ is a Shannon entropy of an $i$th written page calculated in \eqref{eq:entropy} and $N$ is the number of write accesses in $T_{window}$. 

\begin{figure}
\centerline{\includegraphics[width=6.7cm]{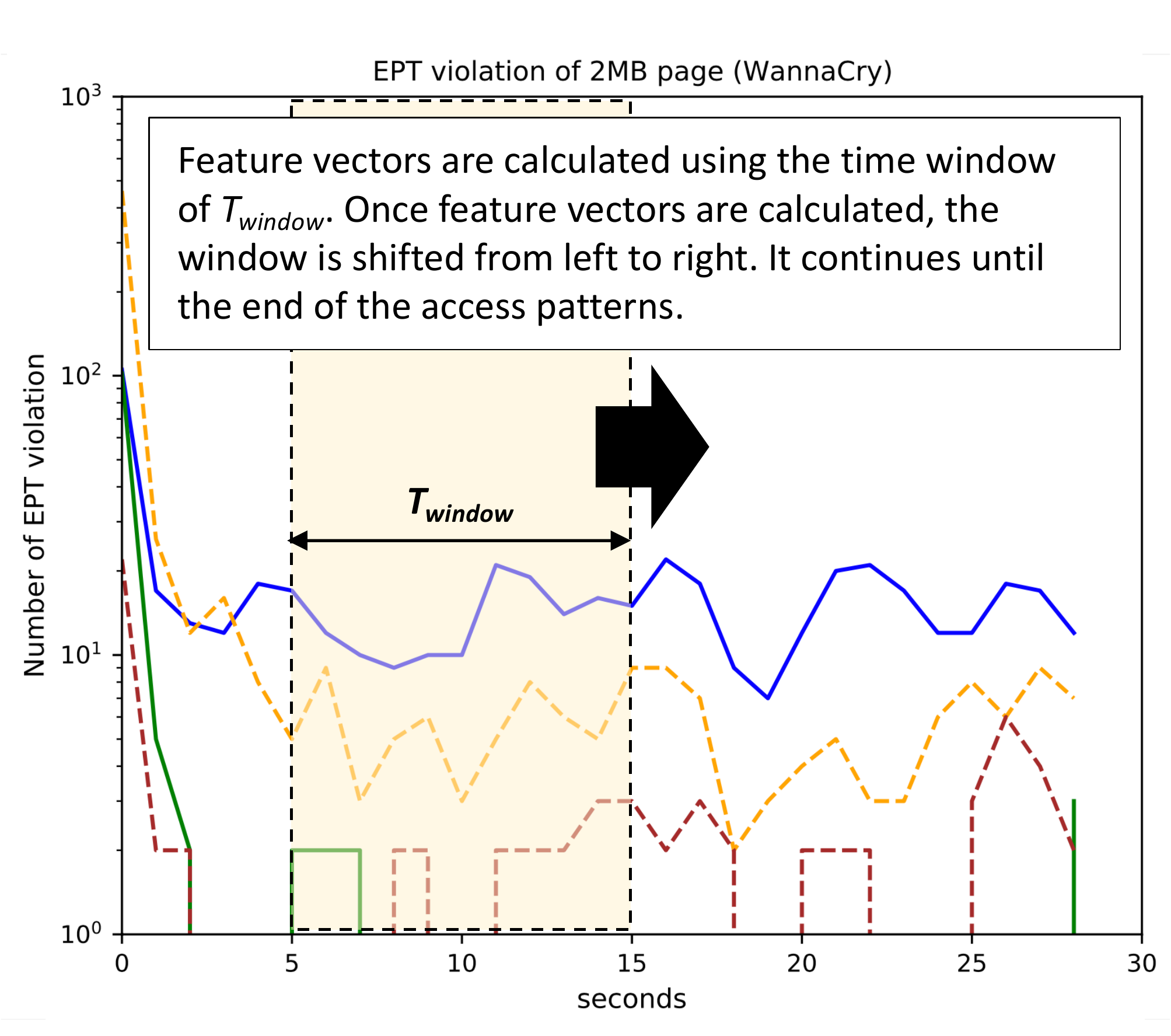}}
\caption{Window-based feature calculation.}
\label{fig:window}
\end{figure}

%% Description of T_{window}
Our feature extractor transforms a set of data points within a time window $T_{window}$ into a single 18-dimensional vector. Fig. \ref{fig:window} show how we calculate the 18-dimensional feature vectors with a window size of $T_{window}$. The feature vectors are calculated by shifting a time window from left to right in time-series memory access patterns. $T_{d}$ is the duration of memory access patterns used for creating a machine learning model. Our feature extractor shifts the time window of $T_{window}$ from $t$ = 0 until $t$ reaches at $(T_d - T_{window})$. When we create feature vectors with $T_{d}$ = 30 s and $T_{window}$ = 10 s, the total number of the 18-dimensional feature vectors will be 20; the model can detect ransomware in 30 s after execution of ransomware. $T_{d}$ is, therefore, referred to as the detection time of ransomware or duration of data points fed into a machine learning model.

$C_{4KB}(t)$, $C_{2MB}(t)$, and $C_{MMIO}(t)$ are the number of accessed pages of 4KB, that of 2MB, and that of MMIO, respectively. A variance of host physical address at $t$ is calculated as follows:

\begin{equation}
    V(t) = \frac{1}{N-1} \sum_{i=1}^{N} ( addr_i - \overline{addr} )^2
\end{equation}

where $\overline{addr}$ is the mean of $addr_i$, $addr_i$ is a physical address of an $i$th access on RAM, $N$ is the number of the accesses in $T_{window}$.

%%%%%%%%%%%%%%%%%%%%%%%%
\section{Dataset creation}

We created a dataset of the eight classes consisting of four malicious classes and four benign classes. Table \ref{tbl:ransomware-hash} shows the hash values of three ransomware samples and one wiper malware sample used in creating the dataset. WannaCry, Sodinokibi (REvil), and Darkside are ransomware samples; CaddyWiper, on the other hand, is a wiper malware sample that destroys data on a victim's computer by wiping all files; and it was discovered in Ukraine \cite{caddywiper}.

We created decoy files from the Govdocs1 dataset that are available in the open repository \cite{govdocs}; the ransomware and wiper malware samples encrypt and delete the decoy files. The Govdocs1 dataset consists of files obtained from U.S. government websites and includes PDF, JPEG, HTML, and Microsoft Office files. We copied the 9,872 files in the first ten directories of the Govdocs1 dataset to the Desktop of a test machine in creating the dataset. 

\begin{table*}[tb]
\begin{center}
\caption{Ransomware and wiper malware samples: SHA-256 hash values and the first submission dates at VirusTotal.}
\label{tbl:ransomware-hash}
\begin{tabular}{|c|c|c|}
\hline
Name & SHA-256 hash & Date \\
\hline
\hline
WannaCry & \texttt{ed01ebfbc9eb5bbea545af4d01bf5f1071661840480439c6e5babe8e080e41aa} & 2017-05\\
Sodinokibi (REvil) & \texttt{0fa207940ea53e2b54a2b769d8ab033a6b2c5e08c78bf4d7dade79849960b54d} & 2019-04\\
Darkside & \texttt{b6855793aebdd821a7f368585335cb132a043d30cb1f8dccceb5d2127ed4b9a4} & 2021-04\\
CaddyWiper & \texttt{a294620543334a721a2ae8eaaf9680a0786f4b9a216d75b55cfd28f39e9430ea} & 2022-03\\
\hline
\end{tabular}
\end{center}
\end{table*}

The dataset includes the four benign classes: Idle, AESCrypt, Zip, and Office suite with a web browser. To create the Idle class dataset, we collected memory and storage access patterns after booting Windows 10 but did not launch any application program. AESCrypt is an open-source symmetric encryption program that uses the Advanced Encryption Standard (AES) algorithm \cite{aescrypt}. We collected memory and storage access patterns when the AESCrypt program (version 3.10) encrypts the 9,872 decoy files. We also collected memory and storage access patterns when the Zip program compresses the directory containing the 9,872 decoy files. The Zip program is a standard compression program included in Windows 10. The memory and storage access patterns of Microsoft Excel, PowePoint, and Firefox were also collected. The version of Excel and PowerPoint is 2019, and that of Firefox is 98.0.2. We created a macro program of Excel that inputs values in cells and creates a graph repeatedly. The slides in PowerPoint are played automatically. Firefox web browser plays a movie on YouTube. We executed the Excel, PowerPoint, and Firefox programs simultaneously in creating the dataset.

Memory and storage access patterns of the above eight classes are collected for 60 s immediately after executing each sample. In every trial, we copied the original contents to the SSD to recover the initial state before ransomware or wiper malware samples destroyed data on the SSD. We collected the access patterns of five trials for each class: the dataset thus contains access patterns of 40 trials in total. 

%%%%%%%%%%%%%%%%%%%%%%%%
\section{Interval for flushing TLBs and clearing EPT}
\label{sec:interval}

\begin{figure}[tb]
    \centering
    \includegraphics[scale=0.5]{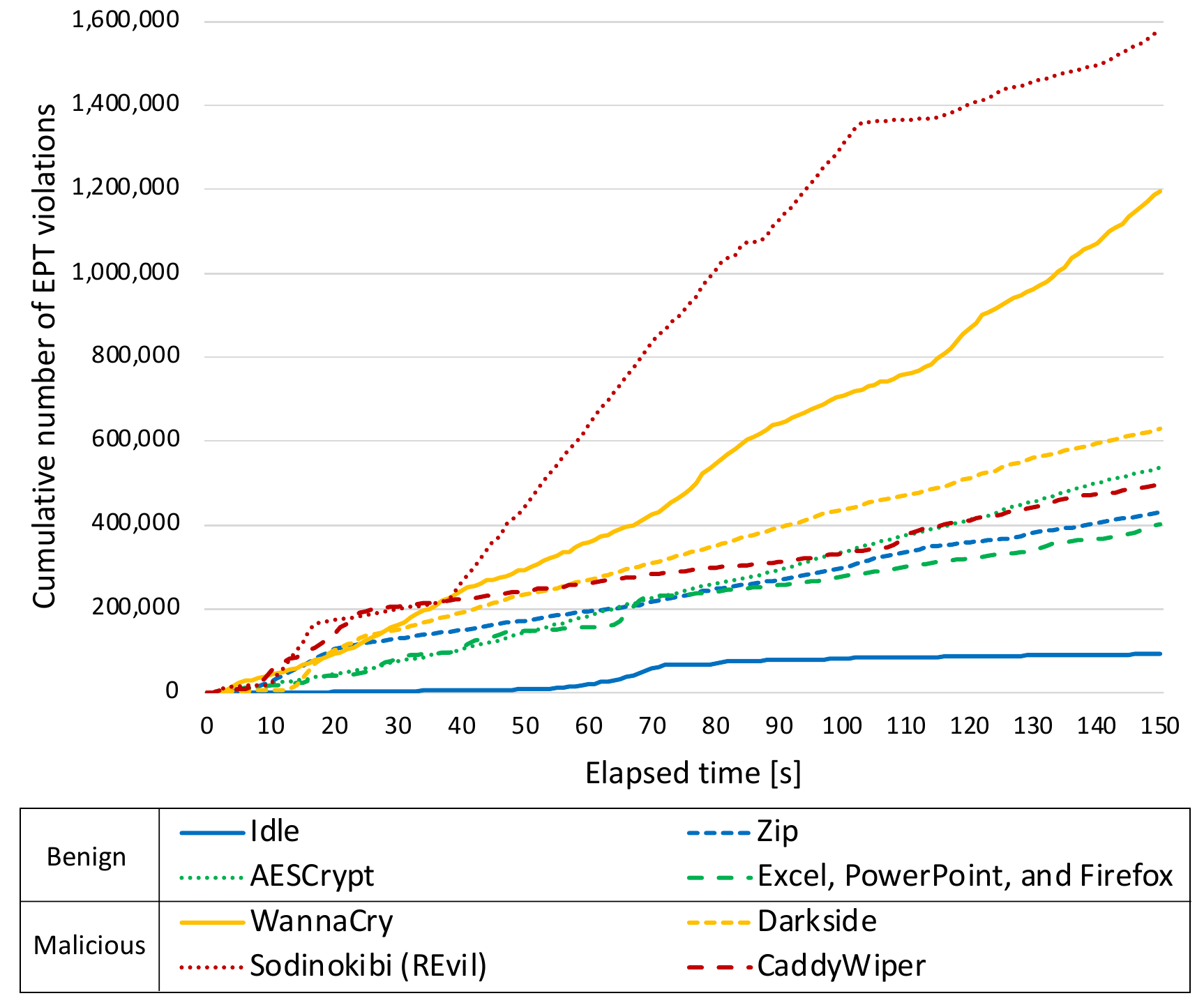}
    \caption{Cumulative number of EPT violations.}
    \label{fig:setup-violation-count}
\end{figure}

Since the developed hypervisor collects memory access patterns using EPT violations, we must determine the appropriate time interval for flushing TLBs on all CPU cores (i.e., TLB shootdown) and deleting all EPT entries. If the hypervisor does not flush TLBs and delete EPT entries at a specific interval, we cannot observe memory accesses except for the first access on each page. Fig. \ref{fig:setup-violation-count} shows the cumulative number of EPT violations immediately after a hypervisor flushed the TLBs on all CPU cores and deleted all EPT entries. The experiment was conducted using the hardware and software shown in Table \ref{tbl:spec-test-machine}. The frequent operations to flush TLBs and to clear EPT entries, in general, cause significant performance degradation of the guest OS. We confirmed that the hypervisor could obtain a sufficient number of EPT violations to monitor memory access patterns in the first 30 s from the experimental result shown in Fig. \ref{fig:setup-violation-count}. In this paper, we determined the time interval of 30 s (i.e., $T_{flush}$ = 30). We, however, need to search for a more appropriate value of $T_{flush}$ through experiments that can achieve both minimizing performance degradation and maximizing the number of observed access patterns.

%%%%%%%%%%%%%%%%%%%%%%%%
\section{Experimental setup}

%% Hyper-parameters
The Random Forest model was trained using the \texttt{RandomForestClassifier} class with the following parameters: the number of trees is 10, and the maximum depth of the tree is 10. The Support Vector Machine (SVM) model was trained using the \texttt{OneVsRestClassifier} class with the radial basis function kernel. The k-Nearest Neighbors (kNN) model was trained using the \texttt{KNeighborsClassifier} class with the number of neighbors $k$ = 5. Scikit-learn 0.23.2 was used in the experiments. We used 1,491 feature vectors of memory and storage access patterns in the first 60 s immediately after executing a sample (i.e., ${T_d}$ = 60 s) with $T_{window}$ = 10 s. 

To evaluate the effectiveness of the models, we used the unweighted mean of $F_1$ scores under 10-fold cross-validation. $F_1$ score is a harmonic mean of precision and recall. An unweighted mean of $F_1$ score is calculated as follows:

\begin{equation}
    \label{eq:f1}
    \overline{F_{1}} = \frac{1}{N} \sum_{i=1}^{N} \left( 2 \cdot \frac{Precision_i \cdot Recall_i}{Precision_i + Recall_i} \right)
\end{equation}

where $N$ is the number of classes. Precision and recall were calculated as follows:

\begin{equation}
    \label{eq:precision}
    Precision_i = \frac{TP_i}{TP_i + FP_i}
\end{equation}

\begin{equation}
    \label{eq:recall}
    Recall_i = \frac{TP_i}{TP_i + FN_i}
\end{equation}

where $TP_i$ (True positive) is the number of correctly classified features belonging to an $i$th class. $FP_i$ (False positive) is the number of incorrectly classified features belonging to an $i$th class. $FN_i$ (False negative) is the number of incorrectly classified features as not belonging to an $i$th class.

%%%%%%%%%%%%%%%%%%%%%%

\section{Classification performance on machine learning models}

\begin{figure*}[tb]
    \centering
    \includegraphics[scale=0.75]{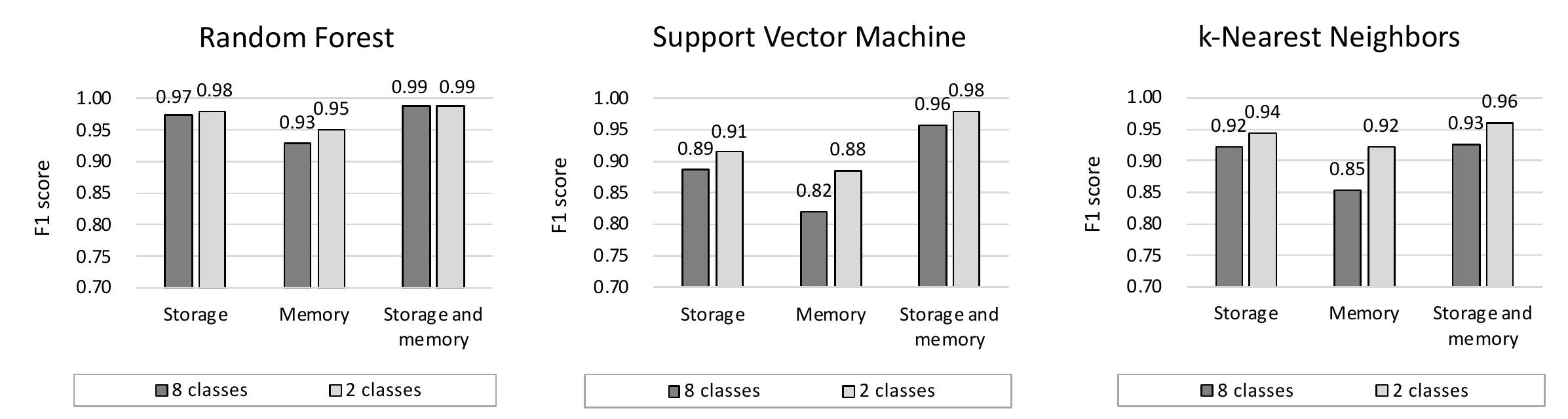}
    \caption{$F_1$ score of the three machine learning models (10-fold cross-validation).}
    \label{fig:ata-ept-all-win_10s-60s}
\end{figure*}

Fig. \ref{fig:ata-ept-all-win_10s-60s} shows the $F_1$ scores of the three machine learning models: Random Forest, SVM, and kNN. The Random Forest model was the best classifier both in 8 and 2 classes. The two-class problem discriminates the four malicious classes (i.e., three ransomware samples and one wiper malware sample) from the four benign samples. Although the $F_1$ scores of memory access patterns are lower than those of storage access patterns, the total $F_1$ scores that used both memory and storage access patterns were improved in all the three models.

\begin{figure}[tb]
    \centering
    \includegraphics[scale=0.60]{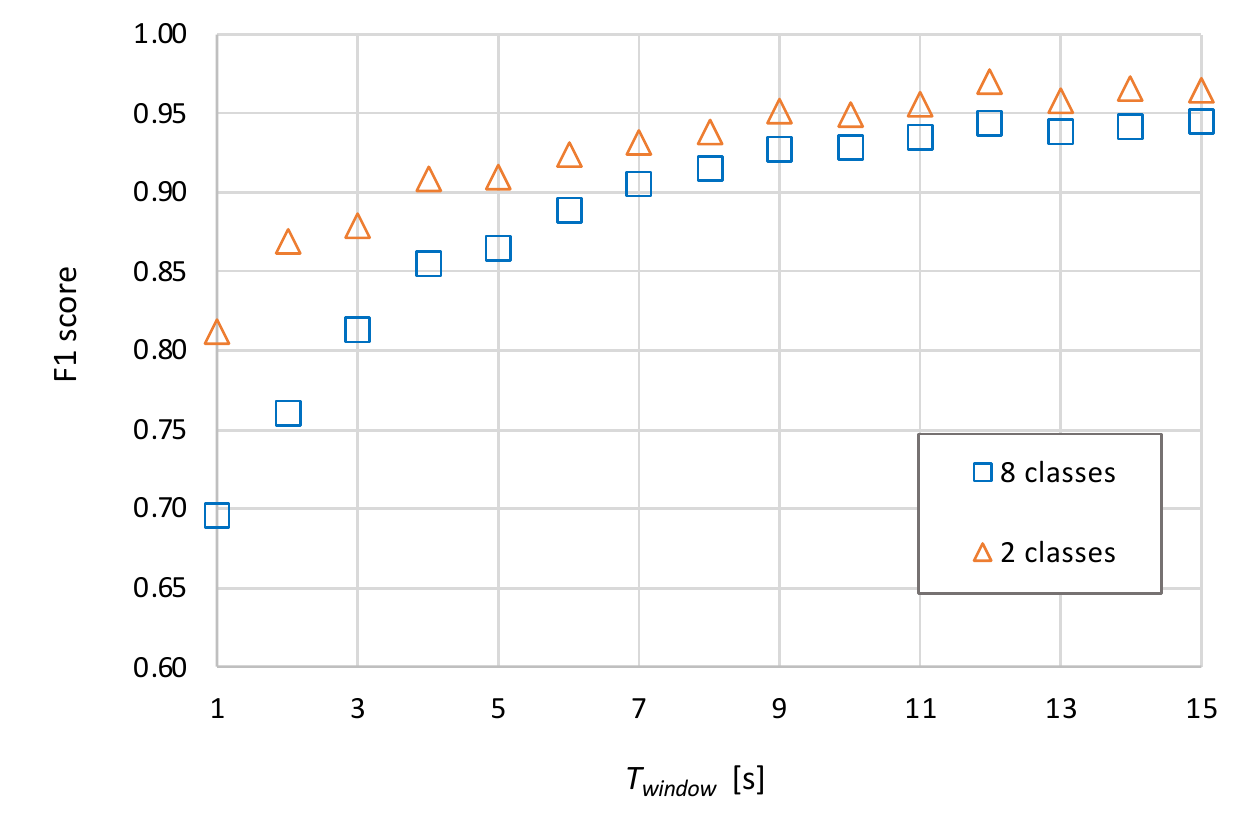}
    \caption{Changes of $F_1$ scores with $T_{window}$ (10-fold cross-validation).}
    \label{fig:time-window-ept-60s}
\end{figure}

%% Time window
We also need to determine the proper time window size, $T_{window}$. Fig. \ref{fig:time-window-ept-60s} shows how $F_1$ scores changed with $T_{window}$ on the Random Forest model that  was trained using memory access patterns with ${T_d}$ = 60 s. The $F_1$ scores saturated at $T_{window}$ above 10; we, therefore, employed $T_{window}$ of 10 s in training the machine learning models.

%%%%%%%%%%%%%%%%%%%%%%%%%
\section{Performance benchmarking}

\begin{figure}
    \centering
    \includegraphics[scale=0.42]{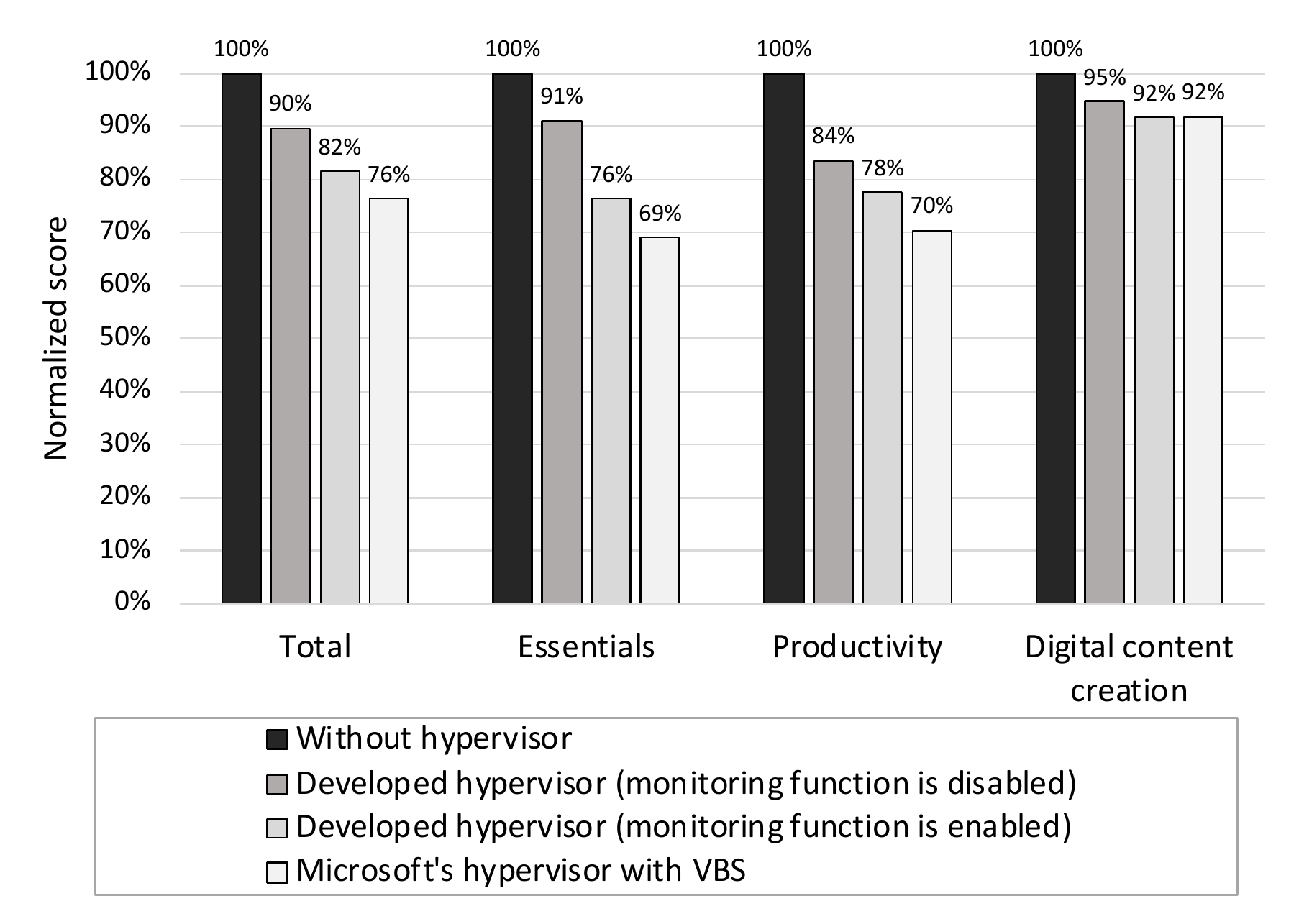}
    \caption{Results of PCMARK10 benchmark tests.}
    \label{fig:PCMARK10}
\end{figure}

Fig. \ref{fig:PCMARK10} shows the results of the PCMARK10 (version 2.1.2525) benchmark that is designed to test typical home user workloads \cite{pcmark10}. For example, the ``Essentials'' workload includes application start-up and web browsing, the ``Productivity'' workload includes spreadsheet and writing, and the ``Digital content creation'' workload includes rendering and video editing.
We first created the baseline of our performance evaluation by measuring a score on the test machine shown in Table \ref{tbl:spec-test-machine} without a hypervisor. We then test the following three systems: (1) the developed hypervisor; the monitoring function is disabled, (2) the developed hypervisor; monitoring function both for memory and storage is enabled, and (3) Microsoft's hypervisor that enabled Virtualization Based Security (VBS) with Hypervisor-protected Code Integrity (HVCI) and credential guard function. Both the developed hypervisor and Microsoft's hypervisor \cite{2021windowsinternals} are Type-1 (i.e., bare-metal) hypervisors. 

Both the developed hypervisor and Microsoft's hypervisor caused performance degradation of 18 \% and 24 \% in total score, respectively. BitVisor, a lightweight and thin hypervisor for research purposes that we employed in the development, enforces security functions using a single guest OS model. Microsoft's hypervisor, on the other hand, enforces more complex security functions using a multiple guest OSs model. For example, when a guest OS on a Windows hypervisor loads a kernel driver, the HVCI function of a privileged guest OS is called to verify the code signature of the kernel driver. The kernel driver is loaded on the guest OS only when verification of the code signature is succeeded. Thus, Microsoft's VBS needs at least two guest OSs simultaneously to enforce security functions. BitVisor, on the other hand, enforces security functions directly from hypervisor software without using any privileged guest OS. This single guest OS model of BitVisor reduces the number of context switching between a guest OS and the hypervisor software and contributes to performance improvement.

%%%%%%%%%%%%%%%%%%%%%%%%%
\section{Visualization of feature vectors}

\begin{figure*}[tb]
    \centering
    \includegraphics[scale=0.7]{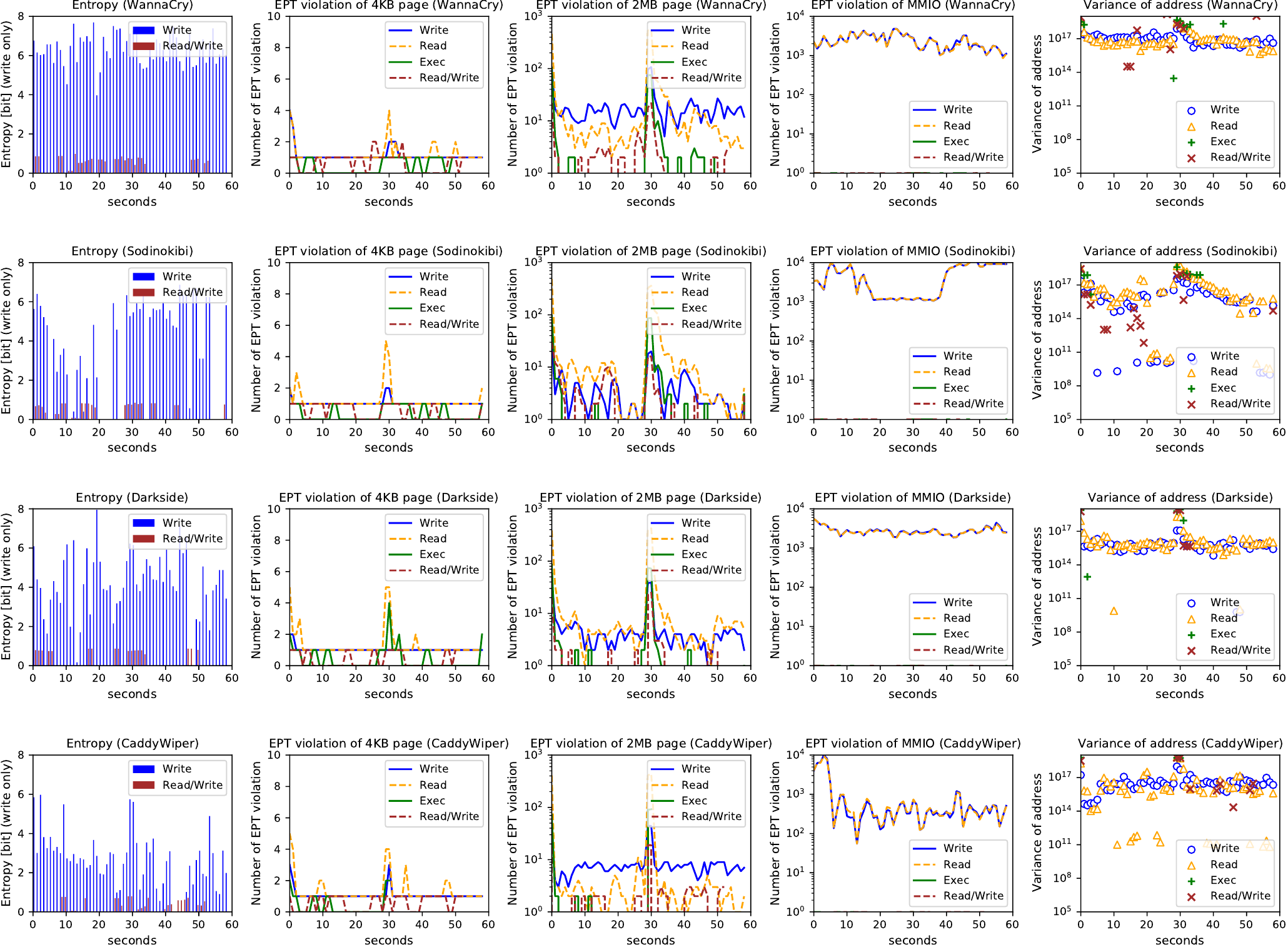}
    \caption{Feature vectors of ransomware and wiper malware samples ($T_{window}$ = 1 s and $T_d$ = 60 s).}
    \label{fig:ransom}
\end{figure*}

\begin{figure*}[tb]
    \centering
    \includegraphics[scale=0.7]{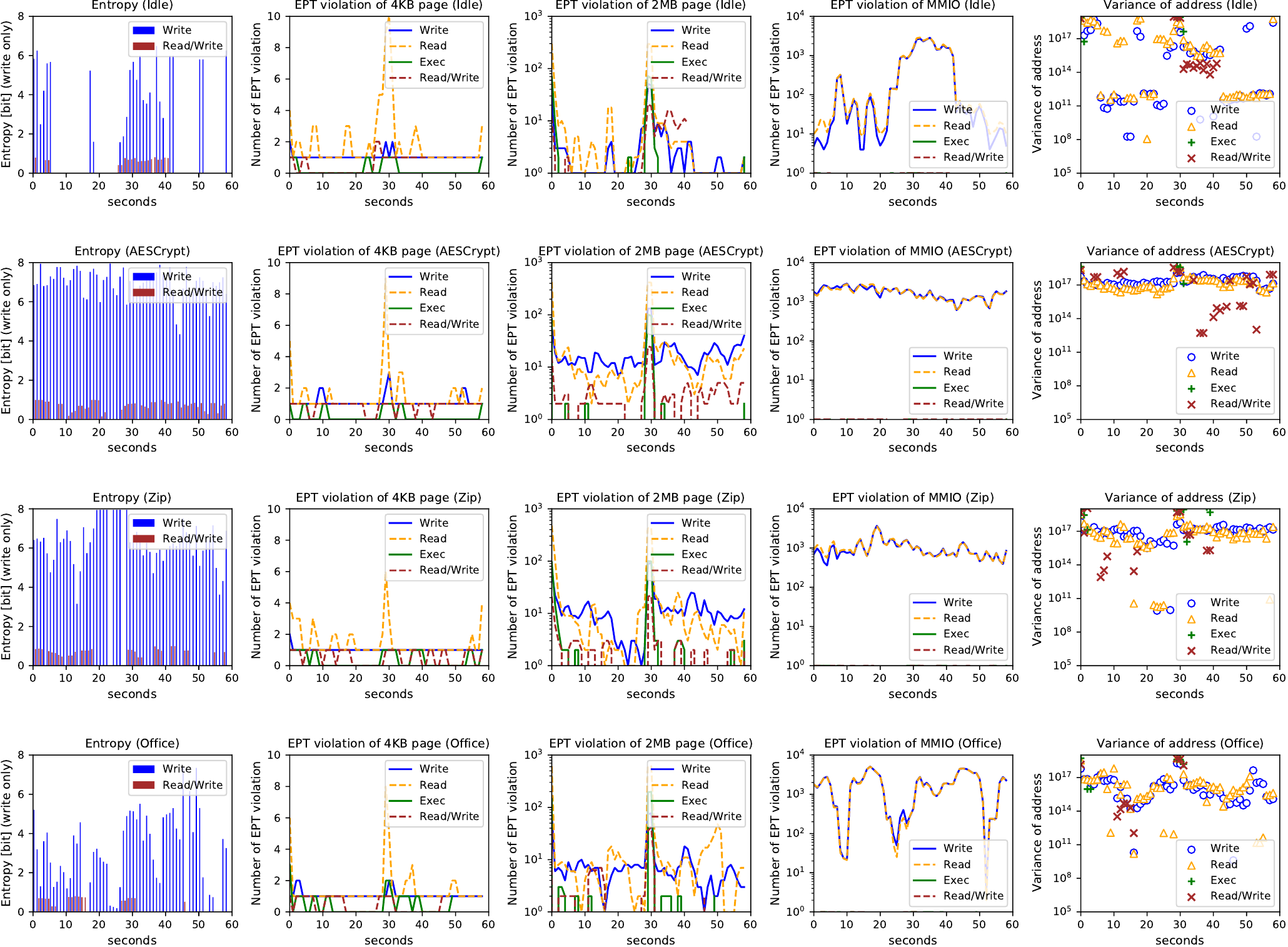}
    \caption{Feature vectors of benign application samples ($T_{window}$ = 1 s and $T_d$ = 60 s).}
    \label{fig:benign}
\end{figure*}

Fig. \ref{fig:ransom} shows feature vectors of ransomware and wiper malware. The entropy of CaddyWiper was lower than those of three ransomware samples because CaddyWiper overwrites files with zeros while ransomware samples encrypt files \cite{caddywiper}. However, both ransomware and wiper malware share common behavioral patterns that heavily access many user files in a short time. Fig. \ref{fig:benign} shows feature vectors of benign application samples. The entropy values of AESCrypt and Zip were higher than those of Idle and Office; The compression algorithm of the Zip program increases entropy because it converts redundant bit patterns into less redundant ones. In the Idle class, we observed some background activities of system programs such as SearchIndexer.exe. Since the hypervisor flushes TLBs and deletes all EPT entries every 30 s, we observed spikes around 30 s on the number of EPT violations of 4KB and 2MB.

%%%%%%%%%%%%%%%%%%%%%%%%%%%%
\section{Conclusion}

We enhanced the functionality of a thin and lightweight hypervisor by adding a function to collect low-level memory access patterns with mitigating advanced evasion techniques of malware that exploit OS vulnerabilities. Furthermore, we presented a design of novel feature vectors of memory access patterns. Our best machine learning classifiers achieved $F_1$ scores of 0.93 in classifying the eight classes and 0.95 in detecting ransomware and wiper malware. Our empirical results show that the developed hypervisor can detect ransomware and wiper malware using only low-level memory access patterns. The current dataset, however, does not include memory access patterns on various conditions, such as the different specifications of chipset, CPU, and RAM. We need to examine the memory access patterns in more realistic situations, for example, when multiple samples are executed simultaneously.

%%%%%%%%%%%%%%%%%%%%%%%%%%%%
\section*{Acknowledgment}

This work was supported by JSPS KAKENHI Grant Number JP20K11825. The authors gratefully acknowledge constructive comments by the anonymous reviewers. The authors thank Takamichi Omori, Keisuke Makihara, and Hiroki Mizuno for their support in developing the hypervisor-based monitoring system. The authors gratefully thank the developers of BitVisor.

\bibliographystyle{IEEEtran}
\bibliography{hirano}

%% Unless there are six authors or more give all authors' names; do not use ``et al.''. 
%% Capitalize only the first word in a paper title, except for proper nouns and element symbols.

\end{document}